\documentclass[useAMS,usenatbib]{mn2e}
\newif\ifAMStwofonts

\def\pmb#1{\mbox{\boldmath$#1$}}
\def\gtsim {>\kern-1.2em\lower1.1ex\hbox{$\sim$}}
\def\ltsim {<\kern-1.2em\lower1.1ex\hbox{$\sim$}}
\def\gtsim {>\kern-1.2em\lower1.1ex\hbox{$\sim$}}
\def\ltsim {<\kern-1.2em\lower1.1ex\hbox{$\sim$}}

\def\be{\begin{equation}}
\def\ee{\end{equation}}

\usepackage{epsfig}
\begin{document}

\title{Excitation of a nonradial mode in a millisecond X-ray pulsar XTE J1751-305}
\author[U. Lee]{
 Umin Lee$^1$\thanks{E-mail: lee@astr.tohoku.ac.jp}
\\$^1$Astronomical Institute, Tohoku University, Sendai, Miyagi 980-8578, Japan}

\date{Typeset \today ; Received / Accepted}
\maketitle


\begin{abstract}
We discuss possible candidates for non-radial modes 
excited in a mass accreting and rapidly rotating neutron star
to explain the coherent frequency identified in the light curves of a millisecond X-ray pulsar XTE J1751-305. The spin frequency of the pulsar is $\nu_{\rm spin}\cong435$Hz and the identified coherent
frequency is $\nu_{\rm osc}=0.5727595\times\nu_{\rm spin}$.
Assuming the frequency identified is that observed in the corotating frame of the neutron star,
we examine $r$- and $g$-modes in the 
surface fluid layer of accreting matter composed mostly of helium, and inertial modes and
$r$-modes in the fluid core and toroidal crust modes in the solid crust.
We find that the surface $r$-modes of $l^\prime=m=1$ and 2 excited by $\epsilon$-mechanism due to helium burning in the thin shell can give the frequency ratio $\kappa=\nu_{\rm osc}/\nu_{\rm spin}\simeq0.57$ at $\nu_{\rm spin}=435$Hz, where $m$ is the azimuthal wave number of the modes.
As another candidate for the observed ratio $\kappa$, we suggest a toroidal crustal mode that has penetrating amplitudes in the fluid core and is destabilized by gravitational wave emission.

Since the surface fluid layer is separated from the fluid core
by a solid crust, the amplitudes of an $r$-mode in the core, which is destabilized by emitting gravitational waves, can be by a large factor different from those in the fluid ocean. 
We find that the amplification factor defined as $f_{\rm amp}=\alpha_{\rm surface}/\alpha_{\rm core}$
is as large as $f_{\rm amp}\sim 10^2$ for the $l^\prime=m=2$ $r$-mode at $\nu_{\rm spin}=435$Hz
for a typical $M=1.4M_\odot$ neutron star model, where $\alpha$'s are
the parameters representing the $r$-mode amplitudes, and $l^\prime$ is the harmonic degree of the mode.
Because of this significant amplification of the $r$-mode amplitudes in the surface fluid layer, 
we suggest that, when proper corrections to the $r$-mode frequency such as due to the general relativistic effects are taken into consideration, the core $r$-mode of $l^\prime=m=2$ can be
a candidate for the detected frequency, without leading to serious contradictions to, for example,
the spin evolution of the underlying neutron star.
\end{abstract}

\begin{keywords}
stars: oscillations -- stars : rotation
\end{keywords}

\section{Introduction}

A recent report of the detection of a coherent frequency from a mass accreting millisecond 
X-ray pulsar XTE J1751-305 (Strohmayer \& Mahmoodifar 2014) suggests the existence of a nonradial mode
excited in the neutron star. 
The spin frequency of the pulsar is $\nu_{\rm spin}\cong 435$Hz and the identified frequency is
$\nu_{\rm osc}=0.5727595\times\nu_{\rm spin}=249.332609$Hz.
If the frequency is really associated with a non-radial mode of a neutron star,
we may be able to rule out $p$-modes for the frequency,
since their oscillation frequencies are higher than kHz in the case of neutron stars
and are too high to be consistent with the detected frequency.
We may also rule out the $g$-modes residing in the core, since they usually have much lower frequencies than the spin frequency of the star because of nearly isentropic structure of the core  
(e.g., McDermott et al 1988).
Therefore, possible candidates remained for the detected frequency will be
a $g$-mode or a rotational mode propagating 
in the surface fluid layer, or a rotational mode in the fluid core, or a toroidal crust mode
in the solid crust.
Note that low frequency $g$-modes and crust modes can be strongly modified by the rapid rotation of
the star.

Accretion powered millisecond pulsars show small amplitude X-ray oscillations with
periods equal to their spin periods, which are assumed to be produced by a hot spot on
the surface of the star (e.g., Lamb et al 2009).
Numata \& Lee (2010) suggested that global oscillations of neutron stars can periodically perturb
such a hot spot so that the oscillation mode periods could be observable as X-ray flux
oscillations.
They also suggested that since the hot spot on the neutron star surface is corotating with
the star, the oscillation frequencies should be equal to those observed
in the corotating frame of the star.

In this paper, we pursue the possibility that the detected frequency in the pulsar is caused by
an unstable non-radial mode of the rapidly rotating neutron star. 
To obtain the oscillation frequency $\omega$ of pulsationally unstable non-radial modes of neutron stars, we calculate the surface $r$-modes and $g$-modes excited by nuclear helium burning in the surface layer for $|m|=1$ and 2, and toroidal crust modes in the solid crust and
rotational modes such as inertial modes and $r$-modes in the fluid core for $m=2$, where
$m$ is the azimuthal wave number of the modes.
Here, $\omega$ denotes the frequency observed in
the corotating frame of the star and is given by
$\omega=\sigma+m\Omega$, where $\sigma$ is the oscillation frequency 
in an inertial frame and
$\Omega=2\pi\nu_{\rm spin}$ is the angular spin frequency of the star.
To calculate surface $r$-modes and $g$-modes, we construct mass accreting and nuclear burning thin 
shells in steady state.
On the other hand, we use a 
neutron star model composed of a surface fluid ocean, a solid crust, and a fluid core,
to compute crust modes in the solid crust and rotational modes in the fluid core.
Note that the crust mode and the core $r$-mode
are expected to be destabilized by emitting gravitational waves.
Assuming $\nu_{\rm osc}=\omega/2\pi$ and $\nu_{\rm spin}=\Omega/2\pi$, we look for non-radial oscillation modes that are pulsationally unstable and
give the ratio $\kappa\equiv\omega/\Omega\simeq0.57$ at $\nu_{\rm spin}=435$Hz.

\section{numerical results}
\subsection{Surface $r$-modes and $g$-modes}

\begin{figure*}
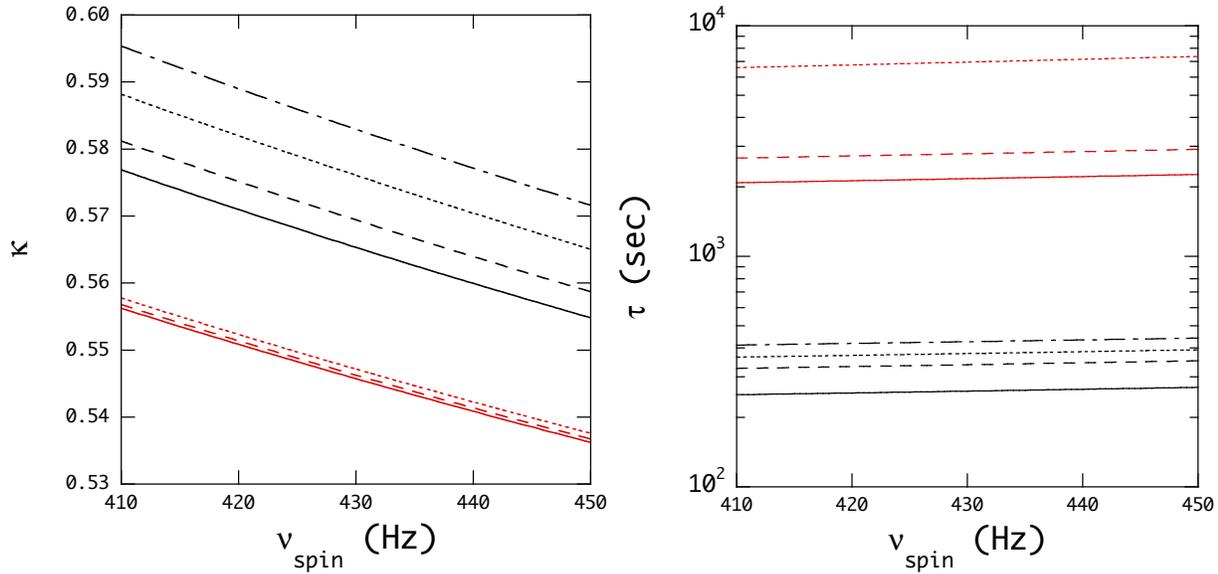

\resizebox{0.45\textwidth}{!}{
\includegraphics{f1a.epsi}}
\resizebox{0.45\textwidth}{!}{
\includegraphics{f1b.epsi}}
\caption{Frequency ratio $\kappa\equiv\omega/\Omega$ and the growth timescale $\tau$ in second
as functions of the spin frequency $\nu_{\rm spin}=\Omega/2\pi$ for the $l^\prime=|m|=1$
$r$-modes propagating in the mass accreting surface fluid shell, where steady burning of hydrogen and helium
is assumed to take place in the shell for the mass accretion rates $\dot M=0.7\dot M_{\rm Edd}$
(black curves) and 0.1 (red curves).
The solid, dashed, dotted, and dash-dotted lines are respectively
for the cases of the hydrogen abundance $X=0,~0.01, ~0.02$, and 0.03 in the accreting matter where $Z=0.02$. }
\end{figure*}

\begin{figure*}
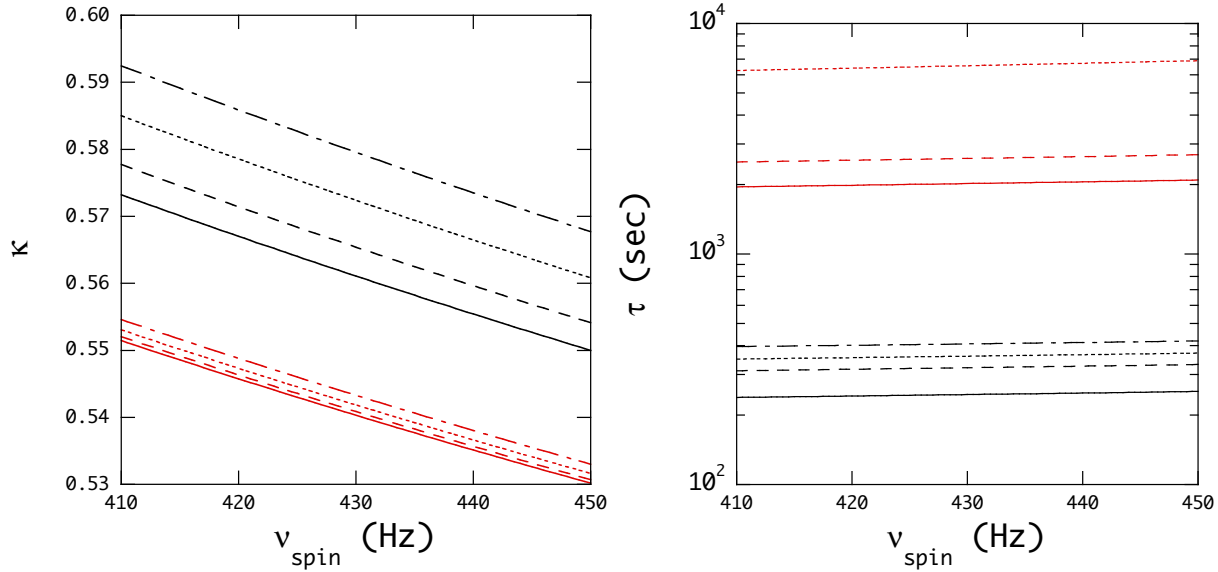

\resizebox{0.45\textwidth}{!}{
\includegraphics{f2a.epsi}}
\resizebox{0.45\textwidth}{!}{
\includegraphics{f2b.epsi}}
\caption{Same as Figure 1 but for $m=2$. }
\end{figure*}

The method of calculation used for $r$- and $g$-modes in the surface fluid ocean is the same as that
given in Strohmayer \& Lee (1996) and Lee (2004).
Following Strohmayer \& Lee (1996), 
assuming steady nuclear burning of hydrogen and helium for a given mass accretion rate $\dot M$,
we compute the surface fluid shell of mass $10^{-10}M_\odot$, which may be regarded as the outermost layer of a mass accreting neutron star.
We assume that the surface layer is radiative, and that accreting matter is 
composed mostly of helium with a small fractional mixture of hydrogen.
We obtain the temperature distributions in the shell similar to those calculated by Lee (2004).
In the region where helium burning takes place, the mean molecular weight rapidly changes with depth, which yields a bump in the Brunt-V\"ais\"al\"a frequency $N$.
The existence of a small fraction of hydrogen in accreting matter slightly enhances the bump, particularly for high mass accretion rates.
When the mass accretion rate is low, the burning layers of helium and hydrogen are well
separated in the shell.
The pulsational stability of oscillation modes propagating
in the thin surface layer
is examined by non-adiabatic oscillation calculation assuming uniform rotation
as described by Lee \& Saio (1987).
For example, the displacement vector $\pmb{\xi}$ in rotating stars
is represented by a finite series expansion in terms of spherical harmonic function $Y_l^m(\theta,\phi)$
for a given $m$:
\be
\xi_r=r\sum_{j=1}^{j_{\rm max}}S_{l_j}(r)Y_{l_j}^m(\theta,\phi)e^{{\rm i}\omega t},
\ee
\be
\xi_\theta=r\sum_{j=1}^{j_{\rm max}}\left[H_{l_j}(r){\partial Y_{l_j}^m(\theta,\phi)\over\partial\theta}
+{T_{l^\prime_j}(r)\over\sin\theta}{\partial Y_{l^\prime_j}^m(\theta,\phi)\over\partial\phi}\right]e^{{\rm i}\omega t},
\ee
\be
\xi_\phi=r\sum_{j=1}^{j_{\rm max}}\left[{H_{l_j}(r)\over\sin\theta}{\partial Y_{l_j}^m(\theta,\phi)\over\partial\phi}
-{T_{l^\prime_j}(r)}{\partial Y_{l^\prime_j}^m(\theta,\phi)\over\partial\theta}\right]e^{{\rm i}\omega t},
\ee
where $l_j=|m|+2(j-1)$ and $l^\prime_j=l_j+1$ for even modes and $l_j=|m|+2j-1$ and
$l^\prime_j=l_j-1$ for odd modes, and $j=1,~\cdots,~j_{\rm max}$.
For even mode, the function $\xi_r$, for example, is symmetric about the equator of the star, while
it is antisymmetric for odd modes.
Note that in this paper no general relativistic effects are considered for the shell 
and mode computations.
For the length of the expansions, we usually take $j_{\rm max}=10$.

The results of non-adiabatic calculation of $r$-modes propagating in the surface thin fluid layer are given in Figures 1 and 2 for $l^\prime=|m|=1$ and 2, respectively, where 
the ratio $\kappa=\omega/\Omega$ and the mode growth timescale $\tau\equiv 1/{\rm Im}(\omega)$
of the $r_1$-mode
are plotted as functions of $\nu_{\rm spin}=\Omega/2\pi$ for 
the mass accretion rates $\dot M/\dot M_{\rm Edd}=0.7$ (black curves) 
and $0.1$ (red curves) for a neutron star model
of the mass $M=1.4M_\odot$ and radius $R=1.179\times 10^6~\rm cm$, where
$\dot M_{\rm Edd}\equiv 4\pi cR/\kappa_e=
1.88\times 10^{18}(1+X)^{-1}(R/10^6)~\rm g~s^{-1}$ is the Eddington mass accretion rate
with $\kappa_e$ being the electron scattering opacity.
Here, the notation $r_1$ indicates that the $r$-mode has one radial node of the eigenfunction.
Note that only $r_1$-mode of $l^\prime=m$ are found pulsationally unstable and
$r$-modes that have radial nodes of the eigenfunctions more than one are all stable.
As discussed by Lee (2004), the frequency $\omega$ of the $r_1$-mode propagating 
in the surface thin shell
becomes insensitive to $\Omega$ for rapid rotation rates and is approximately given by the formula:
\be
\omega\simeq {mN_0(D/R)\over (2j+1)\sqrt{\lambda}},
\label{eq:omegashell}
\ee
where $j$ is an integer associated with the mode, $\lambda$ is the separation constant used to
separate the horizontal motions from the vertical motions in the shell, $N_0$ is the
representative value of
the Brunt-V\"ais\"al\"a frequency in the shell, and $D$ is the depth of the fluid ocean 
(Lee 2004, see also Pedlosky 1987).
We may take $N_0$ as the value at the helium burning layer where the mode excitation takes place.
Because of this insensitiveness of $\omega$ to $\Omega$ the ratio $\kappa$ decreases as 
$\nu_{\rm spin}=\Omega/2\pi$ increases.
The existence of a small amount of hydrogen enhances the value of $N_0$ at the helium burning layer, particularly for high mass accretion rates $\dot M$, and this enhancement leads to an increase
in the ratio $\kappa$ at a given value of $\nu_{\rm spin}$.
The destabilization of the $r$-modes takes place because of the strong temperature dependence
of helium burning in the thin shell.
As indicated by the right panels of Figures 1 and 2, the existence of a small amount of hydrogen 
tends to weaken the destabilization, that is, the mode growth timescale becomes longer
as the hydrogen content is increased.
Because the temperature in the helium burning region in the shell becomes higher
for higher mass accretion rate $\dot M$, the growth timescale $\tau$ becomes
shorter as $\dot M$ increases.

As shown by Figures 1 and 2, if we assume high mass accretion rates $\dot M\sim \dot M_{\rm Edd}$ and 
a small mixture of hydrogen $X$ in the accreting matter, for low $m$ values
we can find pulsationally unstable $r$-modes of $l^\prime=m$ whose
oscillation frequency in the corotating frame is consistent with the ratio $\kappa\simeq0.57$
at $\nu_{\rm spin}\simeq435$Hz.

We have carried out similar calculations for the $r$-modes of $l^\prime=m+1=2$ propagating
in helium burning shells, and 
we have found
that the inertial frame frequency $\sigma=\omega-m\Omega$ of
the $r_1$-mode, which is driven by helium burning, can give the ratio $\hat\kappa\equiv\sigma/\Omega$ consistent with the observed value.

For $g$-modes in the surface helium burning shells, only retrograde $g_1$-modes of $l=m=1$ 
are found to be pulsationally unstable to give the ratio $\hat\kappa$
consistent with the observed value.

\subsection{Crust modes and Core $r$-modes}

It is now well known that the $r$-mode of $l^\prime=|m|=2$ is most strongly destabilized by
gravitational wave radiation (e.g., Andersson 1998; Friedman \& Morsink 1998; Lindblom et al 1998).
For the $l^\prime=|m|$ $r$-mode of entirely fluid stars, the ratio $\kappa=\omega/\Omega$, which 
tends to
$\kappa\rightarrow 2m/l^\prime(l^\prime+1)=2/3$ for $m=2$ as $\Omega\rightarrow 0$, only weakly
depends on $\Omega$ and on the stratification of the fluid (e.g., Yoshida \& Lee 2000).
This fact may suggest that the $r$-mode of $l^\prime=|m|=2$ is unlikely to be responsible for
the observed ratio $\kappa\simeq 0.57$.
But, Andersson et al (2014) recently suggested that a general 
relativistic effect can reduce the ratio $\kappa$ such that the value of $\kappa$ for the $l^\prime=m=2$
$r$-mode becomes consistent with the observed value, depending on the mass $M$, the radius $R$, and the equation of state of the star (e.g., Lockitch et al 2003; see also Yoshida \& Lee 2002).
However, they also argued that the amplitudes of the $r$-mode suggested by the detection of
the coherent frequency is too large to be consistent with
the spin evolution of the star.

The presence of a solid crust in a neutron star, however, makes the modal properties of the star quite complicated.
Because of a solid crust, for example, we have toroidal shear waves propagating in the
solid crust, which affect the $r$-modes and inertial modes in the fluid core.
In fact, using neutron star models with a solid crust, Yoshida \& Lee (2001)
calculated toroidal crust modes and rotational modes ($r$- and inertial-modes) and 
showed that
mode crossings (avoided crossings) between the crust modes and the $r$- and inertial modes
in the core are quite common.
In addition to mode crossing, 
if there is a surface fluid ocean on the solid crust, it is important to note that
there exist $r$-modes propagating in the fluid ocean
besides the $r$-modes in the fluid core (Lee \& Strohmayer 1996; Yoshida \& Lee 2001)
and that the amplitudes of a core $r$-mode penetrate
through the solid crust and are amplified in the surface ocean since the $r$-modes in the 
ocean and in the core have similar frequencies and the mass density in the ocean is much smaller than
in the core.
This means that even if the amplitudes of the $r$-mode is large at the surface so that a surface hot spot
is perturbed with appreciable amplitudes,
the $r$-mode amplitudes in the fluid core can be much smaller than those inferred from the detection of
a coherent frequency in the X-ray light curves.
This reduction of the $r$-mode amplitude in the core will
render the difficulty in the $r$-mode interpretation for the identified frequency
less serious when 
the spin evolution of the star is almost exclusively determined by the core $r$-mode.

The mode crossings mentioned above may be easily understood (Yoshida \& Lee 2001).
In the corotating frame of the star, the oscillation frequency of a toroidal crust mode
may be given by (e.g., Strohmayer 1991)
\be
\omega_{\rm crust}(\Omega)\approx\omega_{\rm crust}(0)+{m\Omega\over l^\prime(l^\prime+1)},
\ee
where $\omega_{\rm crust}(0)$ is the oscillation frequency of the crust mode at $\Omega=0$,
while the oscillation frequency of an $r$-mode may be given by
\be
\omega_{\rm r}(\Omega)\approx{2m\Omega\over l^\prime(l^\prime+1)}.
\ee
For $l^\prime\la 10$, $\omega_{\rm crust}(0)$ of the fundamental toroidal crustal mode 
at $\Omega=0$ is
less than the critical frequency $\Omega_{\rm crit}\equiv\sqrt{GM/R^3}$ and it increases with increasing 
$l^\prime$, while the frequencies $\omega_{\rm crust}(0)$ of overtone modes of the toroidal crustal mode are rather insensitive to
the values of $l^\prime$ (e.g., Lee 2008).
For given $m$ and $l^\prime$, 
since the frequency ratio $\omega_{\rm crust}/\Omega$ can be smaller than $\omega_{\rm r}/\Omega$
for large $\Omega$,
the frequencies of the two modes cross with each other at
\be
\Omega_{\rm cross}\approx{l^\prime(l^\prime+1)\over m}\omega_{\rm crust}(0).
\ee
Because of the mode crossing, which usually results in an avoided crossing, 
the eigenfunctions of the $r$-mode and the toroidal mode
as well as their eigenfrequencies are significantly modified.
If the $l^\prime=|m|=2$ $r$-mode remains unstable because of gravitational wave emission
even around the crossing point, we expect that the crustal modes coupled with the $r$-mode are also destabilized as a result of the coupling.
This suggests a possibility for the existence of an unstable crustal mode with the frequency ratio $\kappa\simeq 0.57$ at the spin rate $\nu_{\rm spin}=435$Hz, although
the mode crossing between the $r$-mode
and the fundamental crust mode for $l^\prime=m=2$ may take place at rather slow rotation rates
(Yoshida \& Lee 2001).

For $m=2$, we calculate $r$-modes, inertial modes and
toroidal crust modes for a $1.4M_\odot$ neutron star model.
The neutron star model, which is composed of 
a surface fluid ocean, a solid crust, and a fluid core, is computed by using a cooling evolution code
of neutron stars, where
the equation of state (EOS) for the core is that by Douchin \& Haensel (2001), EOS for the crust by
Negel \& Vautherin (1973) and Baym, Pethick \& Sutherland (1971), and the surface ocean is assumed to
be made of iron.
We have picked up a neutron star model having the central temperature $T_c\simeq 2\times 10^8$ for
mode computation.
For the solid crust, we assume the average shear modulus $\mu_{\rm crust}=\mu_0$, where
$\mu_0=0.1194(Ze)^2n/a$ and $n$ is the number density of the nuclei and $a$ is the separation between the 
nuclei defined by $4\pi a^3n/3=1$ (Strohmayer et al 1991).
The method of calculation for oscillation modes of the tree component neutron star model
is the same as that in Lee \& Strohmayer (1996), who apply Newtonian dynamics 
in the Cowling approximation, employ the finite series expansions similar to equations (1)$\sim$(3), and
assume adiabatic oscillations.
Using the eigenfunctions of the modes obtained numerically, we compute the mode growth timescale $\tau$  
defined by
${\tau^{-1}}=({2E})^{-1}{dE/ dt}$, where $E$ is the oscillation energy defined by
\be
E={1\over 2}\int\left(\rho\delta v^i\delta v^*_i+{\delta p\over\rho}\delta\rho^*\right)
d^3\pmb{x},
\ee
where $\delta v^i$, $\delta p$, and $\delta\rho$ represent the Eulerian perturbations of
the fluid velocity, the pressure, and the mass desnity, respectively, and the asterisk $(^*)$
implies complex conjugation.
The energy gain rate $dE/dt$ is determined by the sum of various excitation and dissipation mechanisms
(Yoshida \& Lee 2000; see also Ipser \& Lindblom 1991; Lindblom et al 1998), that is, 
\be
{dE\over dt}=\left({dE\over dt}\right)_S
+\left({dE\over dt}\right)_B+\left({dE\over dt}\right)_{GD}+\left({dE\over dt}\right)_{GJ},
\ee
where
\be
\left({dE\over dt}\right)_S=-2\int\eta\delta\sigma^{ij}\delta\sigma^*_{ij}d^3\pmb{x}
\ee
is the dissipation rate due to the shear viscosity with $\eta$ being the
shear viscosity coefficient, and
\be
\left({dE\over dt}\right)_B=-\int\zeta\delta\theta\delta\theta^*d^3\pmb{x}
\ee
is the dissipation rate due to the bulk viscosity with $\zeta$ being the bulk viscosity coefficient, and
\be
\left({dE\over dt}\right)_{GD}=-\sigma\omega\sum_{l=2}^\infty N_l\sigma^{2l}\left|\delta D_{lm}\right|^2
\ee
\be
\left({dE\over dt}\right)_{GJ}=-\sigma\omega\sum_{l=2}^\infty N_l\sigma^{2l}\left|\delta J_{lm}\right|^2
\ee
are the dissipation or excitation rates associated with gravitational wave radiation, 
and the definitions of
the various quantities such as
$\delta\sigma^{ij}$, $\delta\theta$, $\delta D_{lm}$, $\delta J_{lm}$, $N_{l}$, $\eta$, and $\zeta$
are given in Yoshida \& Lee (2000). For the details, see also, e.g., Ipser \& Lindblom (1991), Cutler \& Lindblom (1987), and Sawyer (1989).
Since we are interested in the $r$-modes which are destabilized by emitting 
gravitational waves, we expect that $dE/dt$ and $\tau$ are positive for the modes interested.
Note that we ignore the effects of rotational deformation on the oscillation calculations, and that
no effects of superfluidity in the core on the modal properties and on
the viscosity coefficients are included.

As indicated by equations (12) and (13), gravitational wave emission yields destabilizing contributions to the oscillation modes
with $\sigma\omega<0$, which can be rewritten as
\be
0<\kappa=\omega/\Omega<m,
\ee
and we have $0<\kappa<2$ for $m=2$.
For a mode to be globally unstable due to the gravitational wave emission, however,
the destabilizing contributions need to dominate the sum of all dissipative contributions.

\begin{figure}
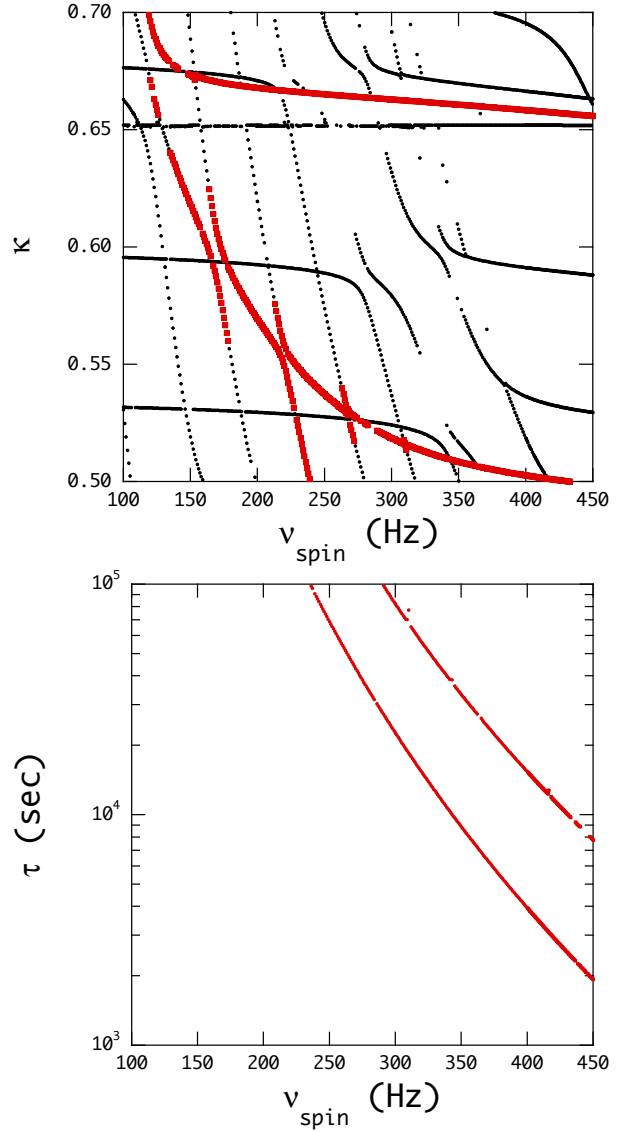

\resizebox{0.45\textwidth}{!}{
\includegraphics{f3a.epsi}}
\resizebox{0.45\textwidth}{!}{
\includegraphics{f3b.epsi}}
\caption{Frequency ratio $\kappa\equiv\omega/\Omega$ and the growth timescale $\tau$ in second
as functions of the spin frequency $\nu_{\rm spin}=\Omega/2\pi$ for the 
$r$-mode of $l^\prime=m=2$ and inertial modes of $m=2$ in the core and toroidal crust modes of $m=2$ in the crust, where the red and black dots indicate unstable and stable modes, respectively, and only unstable modes are plotted for the growth timescale $\tau$.
}
\end{figure}

In Figure 3, we plot the ratio $\kappa$ and the growth time $\tau$ in second
for oscillation modes of the $M=1.4M_\odot$ neutron star model as functions of $\nu_{\rm spin}$, 
where the black and red dots represents stable and unstable modes, respectively, and
$\tau$ is plotted only for unstable modes.
Note that the shear modulus in the crust is set equal to $\mu_{\rm crust}=\mu_0$.
Note also that we have not tried to obtain a complete mode distribution by calculating 
every detail of mode crossings in the $\nu_{\rm spin}$-$\kappa$ plane.
Almost horizontally running black curves, which experience mode crossings with
toroidal crust modes, indicate inertial modes in the core.
Figure 3 shows that mode crossings are quite common between the toroidal crust modes
and inertial modes or $r$-modes, and that the effects of mode coupling between the modes belonging to the same $l^\prime$ are significant so that the crossing
appreciably modifies the frequencies and eigenfunctions, which is particularly true
between the toroidal crustal modes and $r$-mode of $l^\prime=m=2$.
In other words, the crossings between modes of different $l^\prime$'s are 
not necessarily strong enough to significantly
modify the mode properties near the crossing point.
The mode along the red curve, running almost parallel to the line of $\kappa\simeq0.65$
after the crossing at $\nu_{\rm spin}\sim 130$Hz, has the shortest growth timescale $\tau$ 
at every $\nu_{\rm spin}$ and can be regarded as the $r$-mode of
$l^\prime=m=2$.
The eigenfunctions of the mode with $\kappa\cong0.6567$ at $\nu_{\rm spin}=435$Hz are shown in Figure 4, where the expansion coefficients $xiT_{l^\prime_1}$, $xH_{l_1}$, and $xS_{l_1}$ are plotted versus $x\equiv r/R$ and, in the inset, versus $\log(1-x)$, and the amplitude normalization is given by $iT_{l^\prime_1}=1$ at the surface $x=1$.
The toroidal component is totally dominating over the other components both in the fluid regions and in the solid crust, indicating the mode is an $r$-mode.
At the bottom of the solid crust, the amplitude of the toroidal component is only 1\% of the
amplitude at the surface, indicating that the amplification of the mode amplitude occurs
between the fluid core and the surface ocean.

As shown by Figure 3, we find no unstable modes with the ratio $\kappa\simeq0.57$ at
$\nu_{\rm spin}=435$Hz, but we find another unstable mode with $\kappa\cong0.4999$, the eigenfunctions of which
are plotted in Figure 5.
The figure shows that the toroidal component is dominating both in the crust and in the fluid core
although the spheroidal components of the displacement vector also have appreciable amplitudes in both of the regions.
Note also that the amplitude of the eigenfunctions in the surface ocean
is not amplified by a large factor compared with the amplitude in the crust and fluid core, 
indicating that no strong amplification
of the amplitudes takes place between the fluid core and ocean.
The properties of the eigenfunctions suggest that the mode is regarded as a toroidal crust mode and that 
the amplitudes of the mode penetrate into the fluid core as a result of the effects of rapid rotation.
This penetration of the amplitudes into the fluid core makes the mode unstable by emitting gravitational waves.

\begin{figure}
\resizebox{0.45\textwidth}{!}{
\includegraphics{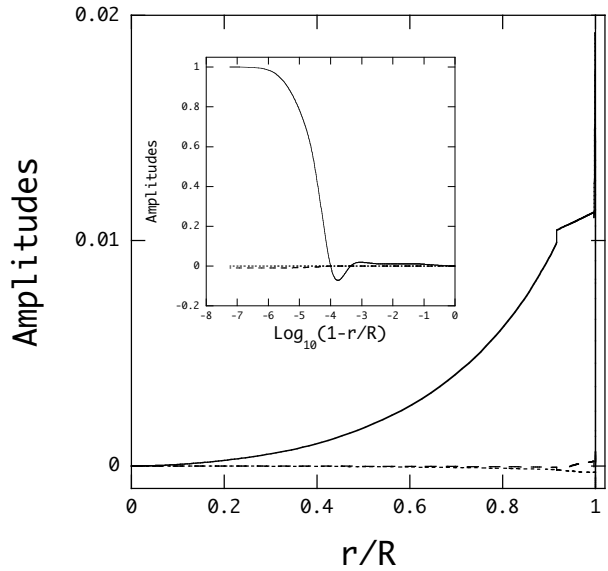}}
\caption{Eigenfunctions of an unstable $l^\prime=m=2$ $r$-mode with the ratio $\kappa\cong0.6567$ at $\nu_{\rm spin}=435$Hz
for the $1.4M_\odot$ neutron star model with the shear modulus $\mu_{\rm crust}=\mu_0$, where the dotted, dashed, and solid lines are for $xS_{l_1}$, $xH_{l_1}$, and $xi T_{l^\prime_1}$, respectively,
and $x=r/R$.
The amplitude normalization is given by $iT_{l^\prime_1}=1$ at the surface of the star. The same eigenfunctions are plotted
versus $\log(1-r/R)$ in the inset.
}
\end{figure}

\begin{figure}
\resizebox{0.45\textwidth}{!}{
\includegraphics{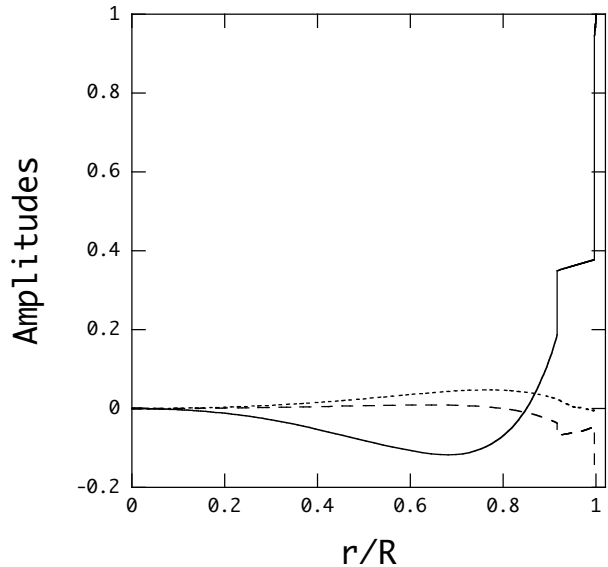}}
\caption{Same as Figure 4 but for an unstable toroidal crust mode of $l^\prime=m=2$ with the ratio $\kappa=0.4999$ at $\nu_{\rm spin}=435$Hz.
}
\end{figure}

Figure 3 shows that for the case of $\mu_{\rm crust}=\mu_0$,
there exist neither unstable crust modes nor unstable $r$-modes which
give the ratio $\kappa\simeq 0.57$ at $\nu_{\rm spin}=435$Hz.
As suggested by Strohmayer \& Mahmoodifar (2014), if we are allowed to increase the crust shear modulus $\mu_{\rm crust}$ to
$\mu_{\rm crust}\simeq 5\times \mu_0$, for example, we do obtain an unstable toroidal crust mode
with $\kappa\simeq 0.57$ at $\nu_{\rm spin}=435$Hz, consistent with the observed ratio.
This ad hoc treatment may not be necessary if we correctly take into account the general relativistic effects on the frequency of the $r$-mode as discussed by Andersson et al (2014).
However, it is still worthwhile to keep in mind the possibility that a toroidal crust mode, which would be
destabilized by emitting gravitational waves, gives the observed ratio $\kappa\simeq 0.57$ at $\nu_{\rm spin}=435$Hz for a certain reasonable combination of physical quantities such as the mass $M$,
the radius $R$, the shear modulus $\mu_{\rm cust}$,
and the equation of state used to construct a neutron star model.

\section{conclusion}

We have discussed candidates of non-radial modes for the detected frequency $\nu_{\rm osc}=
0.5727\times \nu_{\rm spin}$ at $\nu_{\rm spin}=435$Hz found for the millisecond X-ray pulsar
XTE J1751-305.
We have shown that the $r_1$-modes and $g_1$-modes propagating in the surface fluid layer of
accreting matter composed mostly of helium with a small mixture of
hydrogen are pulsationally unstable and can be responsible for the frequency
detected.
We have found that toroidal crustal modes of $l^\prime=m=2$, which have appreciable amplitudes
in the fluid core because of the effects of rapid rotation,
are destabilized by emitting gravitational waves, although the strength of
the destabilization is weaker that that for the $r$-mode of $l^\prime=m=2$.
We have also suggested a possibility that an unstable toroidal crust mode of a neutron star model can be responsible for the observed periodicity in the pulsar.

We have shown that for the $r$-mode of $l^\prime=m=2$
there occurs a strong amplification of the amplitudes
between the fluid core and the surface fluid ocean, an amplification
as large as $f_{\rm amp}\equiv \alpha_{\rm surface}/\alpha_{\rm core}\sim 10^2$, where $\alpha$'s 
are the parameters representing the $r$-mode amplitudes.
As discussed by Strohmayer \& Mahmoodifar (2014), the strength of the detected frequency
indicates the amplitude of $\alpha_{\rm surface}\sim 10^{-3}$, for which 
the $r$-mode amplitudes in the fluid core becomes $\alpha_{\rm core}\sim 10^{-5}$ for $f_{\rm amp}\sim 10^2$.
This significant reduction in the $r$-mode amplitudes in the core will
render much less serious the difficulty met in the $r$-mode interpretation for the detected frequency,
since the spin change rate and heating rate of the star due to the $r$-mode excitation become 
much smaller than those inferred from the detection of the frequency.
Note that, as Andersson et al (2014) discussed, 
if various frequency corrections such as due to the general relativity are taken into account
(see e.g. Yoshida \& Lee 2002; Lockitch, Friedman \&
Andersson 2003),
it is possible to obtain the ratio $\kappa\simeq 0.57$ for the $l^\prime=m=2$ $r$-mode, for which
the ratio tends to $\kappa=2/3$ in the limit of $\Omega\rightarrow 0$ in the Newtonian gravity.
Probably, we need a larger amplification factor $f_{\rm amp}$ to completely
remove the difficulty, since Mahmoodifar \& Strohamyer (2013), for example, suggested the $r$-mode amplitudes ranging from
$\alpha\sim10^{-8}$ to $\sim10^{-6}$.
We need more careful discussions and calculations for the determination of the factor $f_{\rm amp}$ for the $r$-modes of neutron star models with a solid crust.
The factor $f_{\rm amp}$ may depend on the structures of the ocean and the crust.
We need to compute the $r$-mode in the general relativistic frame work with proper treatments
of the jump conditions at the interfaces between the solid crust and the fluid regions.
The existence of a weak magnetic field possibly affects the amplification.

To estimate the effects of the viscous boundary layer on the stability we use an extrapolation formula (Bildsten \& Ushomirsky 2000; Andersson et al 2000; Yoshida \& Lee 2001) given by
\be
{1\over\tau_{\rm VBL}}={1\over \tilde\tau_{\rm VBL}}\left({10^8~{\rm K}\over T_c}\right)
\left({\Omega^2\over \pi G\bar\rho}\right)^{1/4},
\ee
where $\bar\rho=M/(4\pi R^3/3)$.
If we take the value $\tilde\tau_{\rm VBL}=3.7\times10$ (e.g., Yoshida \& Lee 2001), we have $\tau_{\rm VBL}\sim 10^2$ for $T_c\simeq 2\times 10^8$K, 
which is shorter than the growth timescales of the $l^\prime=m=2$ $r$-mode and toroidal crust mode
computed in this paper,
suggesting that these modes are damped by the viscous boundary layer effects.
If the transition between the solid crust and the fluid core is not sharp enough for a
thin viscous boundary layer to form, the effects of viscous dissipations at the boundary
will be weak and probably the destabilized $r$- and crust modes by emitting gravitational waves remain unstable
(e.g., Bondarescu \& Wasserman 2013).

If the frequency detected in the X-ray pulsar XTE J1751-305 is really produced by
a non-radial mode of the underlying neutron star and if it is possible to 
obtain a correct mode identification
for the frequency, 
we will be able to use the mode to probe the physical properties of the star, such as
the mass $M$, the radius $R$, the shear modulus $\mu_{\rm crust}$, and equation of state.
If the frequency is due to a toroidal crustal
mode or an $r$-mode destabilized by emitting gravitational waves, 
the detection of the oscillation frequency can be regarded as an evidence for the existence of a neutron star radiating gravitational waves with detectable amplitudes, which will be useful for
understanding the physics expected in strong gravity environment.

\end{document}